\documentclass[journal]{IEEEtran}

\usepackage{cite}
\usepackage{amsmath,amssymb,amsfonts,bm}
\usepackage{graphicx}
\usepackage{textcomp}
\usepackage{xcolor}
\usepackage{multicol}
\usepackage{booktabs}
\usepackage{multirow}
\usepackage{gensymb}

\usepackage{soul}
\usepackage{graphicx}
\usepackage{amsmath}
\usepackage{booktabs}
\usepackage{threeparttable}
\usepackage{algorithm,algorithmic}

\newcommand*{\etal}{\textit{et al}}
\newcommand*{\vect}[1]{\text{\textbf{#1}}}

\newcommand*{\temp}[1]{#1~{\degree}C}
\newcommand{\pluseq}{\mathrel{+}=}

\begin{document}

\title{Design Exploration and Security Assessment of PUF-on-PUF Implementations}

\author{\IEEEauthorblockN{Kleber Stangherlin, Zhuanhao Wu, Hiren Patel, Manoj Sachdev} \\
        \IEEEauthorblockA{ECE Department, University of Waterloo, Waterloo, ON N2L 3G1, Canada}\\
        \{khstangh, zhuanhao.wu, hiren.patel, msachdev\}@uwaterloo.ca
}

\markboth{Submitted to IEEE for possible publication. Copyright may be transferred. This version may no longer be accessible.}{}

\maketitle

\begin{abstract}
We design, implement, and assess the security of several variations of the PUF-on-PUF (POP) architecture. We perform extensive experiments with deep neural networks (DNNs), showing results that endorse its resilience to learning attacks when using APUFs with 6, or more, stages in the first layer. Compositions using APUFs with 2, and 4 stages are shown vulnerable to DNN attacks. We reflect on such results, extending previous techniques of influential bits to assess stage bias in APUF instances. Our data shows that compositions not always preserve security properties of PUFs, the size of PUFs used plays a crucial role. We implemented a testchip in 65~nm CMOS to obtain accurate measurements of uniformity, uniqueness, and response stability for our POP implementations. Measurement results show that minimum bit error rate is obtained when using APUFs with 8 stages in the first layer, while fewer APUF stages lead to a large spread of bit error rate across different chips.
\end{abstract}


\section{Introduction}

Strong physical unclonable functions (PUFs) offer a solution for the counterfeiting problem of integrated circuits (ICs). They harvest device specific characteristics to generate a digital fingerprint used to authenticate an IC. It is infeasible to manufacture two identical PUFs. Different PUF instances will always produce distinct fingerprints. Strong PUF authentications are performed with a challenge-response protocol~\cite{seminalArb2002}. The mapping from challenges to responses is unique for each PUF. This is unlike traditional authentication methods, where two ICs can be programmed to produce the same proof of identity.

However, researchers have shown that learning attacks are able to predict responses after seeing a limited number of challenge-response pairs (CRPs)~\cite{seminalAttack2013}. A well known counter-measure for learning attacks is to surround the PUF with keyed cryptographic operations such as SHA or AES~\cite{seminalControlled2002}. Although effective in increasing security, such solutions have large area cost, which restricts the range of applications that can benefit from strong PUFs. Constructing low-cost strong PUFs capable of withstand learning attacks is an unsolved problem, and an active field of research.

\begin{figure}[t]
    \centering
    \includegraphics[scale=1]{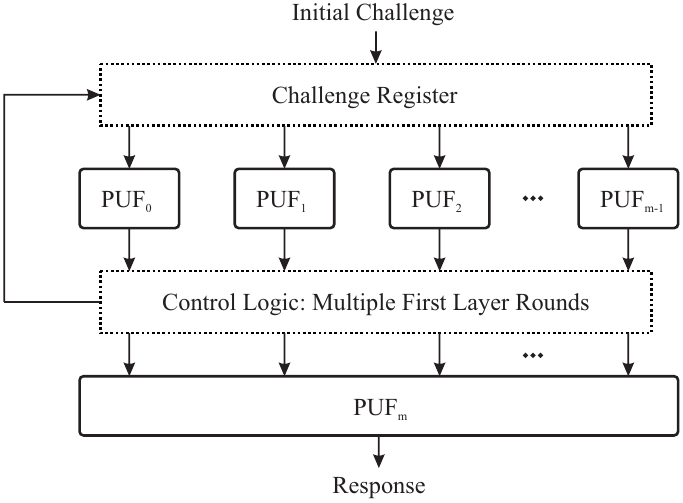}
    \caption{Composite strong PUF architecture supporting multiple first round evaluations. The design extends the PUF-on-PUF (POP) concept, proposed in~\cite{pufPop2019}.}
    \label{fig:arch}
\end{figure}

In this paper, we study the effect of composite architectures in the learning resilience and response stability of PUFs. In particular, we designed and implemented a testchip in 65~nm CMOS with several variations of the PUF-on-PUF (POP) architecture, introduced by Wu \etal~\cite{pufPop2019}. POP uses a two layers construction, shown in Fig. \ref{fig:arch}, where responses from first layer serve as input to the second layer. Our implementation also adds new features to POP, including support for multiple first round evaluations, and temporal majority voting (TMV) for noise removal. We use APUFs as basic building block. Our testchip includes different implementations of the first layer using APUF of 2, 4, 6, 8, 12, and 24 stages.

To assess the security of our POP implementations, we performed extensive learning attacks varying number of rounds and size (number of stages in first layer APUFs). We show that attacks using deep neural networks (DNNs) fail to predict responses when the first layer uses APUFs of 6, 8, 12, and 24 stages. We also found counter-intuitive results for learning resilience. POP implementations using 2, and 4 stage APUFs in the first layer are vulnerable to DNN attacks. Moreover, increasing the number of rounds brought no improvements against DNN attacks, in fact, it made small sized POP implementations more vulnerable. We reflect on such results, extending previous techniques of influential bit analysis to assess stage bias in APUF instances~\cite{attackInfbits2016}. To shed light on why depth increase is ineffective to thwart learning attacks, we show that the hamming-distance of first layer responses decreases as the number of rounds increases. Therefore, small APUFs in the first layer limit the challenge space of the second layer APUF, showing that compositions not always preserve security properties of PUFs --- the size of PUFs used plays a crucial role.

Other contributions of our work include extensive measurements from our testchip, including uniformity, uniqueness, and bit error rate for different implementation sizes. We show that there is an optimal number of stages to improve APUF response stability. Minimum bit error rate is achieved when APUFs with 8 stages are used in the first layer. Further reduction in the number of stages causes a steep increase in the spread of bit error rate across different chips.

This work is organized as follows. First, section \ref{sec:relworks} brings a review of relevant works, and current state-of-the-art in composite strong PUF design. Sections \ref{sec:notation} and \ref{sec:back} cover notation and background, including strong PUF performance metrics, and a system level perspective on the impact of bit error rate. The POP architecture is described in section \ref{sec:comp}, and our testchip design is detailed in section \ref{sec:dsgn}. Measurement results for our testchip are presented in section \ref{sec:meas}. Our strategy for security assessment, as well as our attack results, are presented in section \ref{sec:sec}, including an analysis on probability of output change, influential bits, and hamming distance of intermediate responses. Finally, section \ref{sec:conc} presents the conclusion.

\begin{figure}[t]
    \centering
    \includegraphics[scale=1]{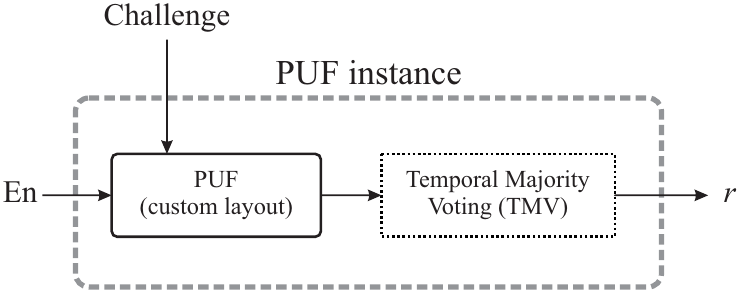}
    \caption{Temporal majority voting (TMV) is implemented in each individual PUF instance. It evaluates the PUF multiple times, returning the most frequent response.}
    \label{fig:tmv}
\end{figure}

\begin{figure}[t]
    \centering
    \includegraphics[scale=1]{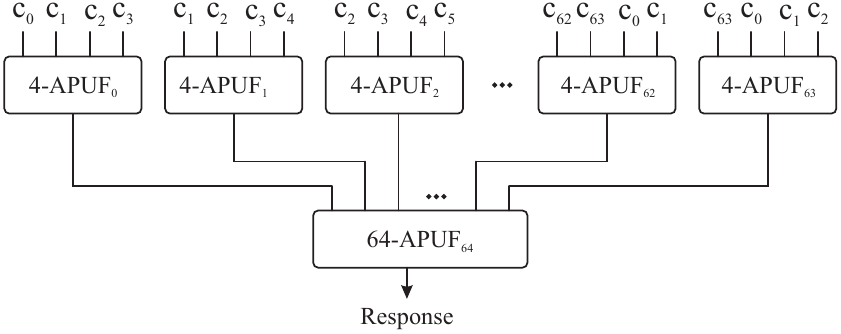}
    \caption{Input challenge wiring in first layer for APUFs of size 4. Other sizes follow similar pattern, where challenge bits are applied to multiple PUF instances to avoid cryptanalysis attacks.}
    \label{fig:wiring}
\end{figure}

\section{Related Works\label{sec:relworks}}

The arbiter PUF (APUF) was introduced as the first silicon PUF~\cite{seminalArb2002}. PUF architectures have been continuously enhanced to improve resilience against learning attacks~\cite{pufFeedforward2004, pufXorAndROPool2007, pufLightweight2008, pufDoubleArb2014, pufMuxPuf2017}. Other works have tried to design PUFs with non-linear challenge-response relationship by operating in subthreshold regime~\cite{pufTaiwan2021, pufSubvt2019}, or using amplifier chains~\cite{pufAmpChain20214}. Such solutions were effective in improving resilience to classical machine learning approaches \cite{seminalAttack2013}, but none was shown resistant to recent attacks using deep neural networks \cite{attackDNN2019, attackMuxPufDnn2019}.

Composite architectures use the output of PUFs as input of other PUFs. They were initially introduced in~\cite{pufCpuf2014}. Later, researchers also proposed combining weak and strong PUFs~\cite{pufMPuf2018}. The concept of composite architectures was used in the interpose PUF, where multiple XOR-APUF instances form a composition with improved resilience to learning attacks~\cite{pufInterpose2019}. Composite constructions require larger CRP datasets, and longer training time, but still, security against DNN attacks remain an unsolved problem~\cite{attackDNN2019, attackMuxPufDnn2019, attackIPUF2021, attackIPUFSplit2020, pufPop2019}.

The work by Wu~\etal~\cite{pufPop2019} highlights the vulnerabilities of prior composite PUFs against cryptanalysis attacks \cite{attackCryptanalysis2015}, and introduces a new architecture which is resistant against such attacks, denoted as PUF-on-PUF (POP). In this work, we explore the design space of POP, implementing a testchip in 65~nm CMOS process, and performing an extensive security assessment on various POP implementations.

\section{Notation\label{sec:notation}}

An arbiter PUF (APUF) with an $n$-bit challenge is denoted as an APUF size $n$, or $n$-APUF. An $n$-APUF-POP refers to a POP implementation where all APUFs in first layer have $n$ stages. Vectors are written in bold text, and are indexed from zero, for example, $\vect{c} = (c_0, c_1, \ldots, c_{n-1})$. The hamming weight and hamming distance functions are denoted as HW(), and HD(), respectively. The \textit{narrow}, and \textit{wide} temperature sets refer to \{\temp{0}, \temp{20}, \temp{60}\}, and \{\temp{-30}, \temp{0}, \temp{20}, \temp{60}, \temp{80}\}.

\section{Background\label{sec:back}}

The essential difference between strong PUFs and conventional authentication methods is the lack of blank samples. Two authentic ICs using PUFs will never produce the same digital fingerprint, regardless of their programmed memory content. The challenge response protocol used to authenticate PUFs still requires the programming of a chip identifier, which is used to fetch the corresponding CRP database enrolled for that PUF instance. Authentication is performed by inquiring the PUF with a subset of the challenges in the database. To avoid replay attacks, challenges are never used more than once. If the number of correct responses exceeds an application defined threshold, the IC is deemed authentic.

\subsection{Performance Metrics\label{sec:metrics}}

The quality assessment of strong PUFs uses metrics that evaluate uniformity, uniqueness, and stability of responses.

\subsubsection{Uniformity} estimates the ratio of zeros and ones in PUF responses. It is also known as ``normalized hamming weight". Ideal uniformity is 0.5, which indicates, on average, equal number of zeros and ones.

\subsubsection{Uniqueness\label{sec:bg:uniq}} estimates the distance between responses from multiple instances. It is also known as ``normalized hamming distance". Ideal uniqueness is 0.5, which indicates that, for the same set of challenges, on average, half responses will differ.

\subsubsection{Bit error rate (BER)} estimates reproducibility of responses under several environmental conditions. Bit error rate (BER) reports a ratio of bits (responses) that differ from their enrolled value. BER ideal value is 0\%, which indicates no incorrect responses during measurement. Other literature may use the term reliability, which simply denotes (100\% - BER).

\subsection{Authenticating with Non-ideal Bit Error Rate\label{sec:auth}}

\begin{figure}[t]
    \centering
    \includegraphics[scale=1]{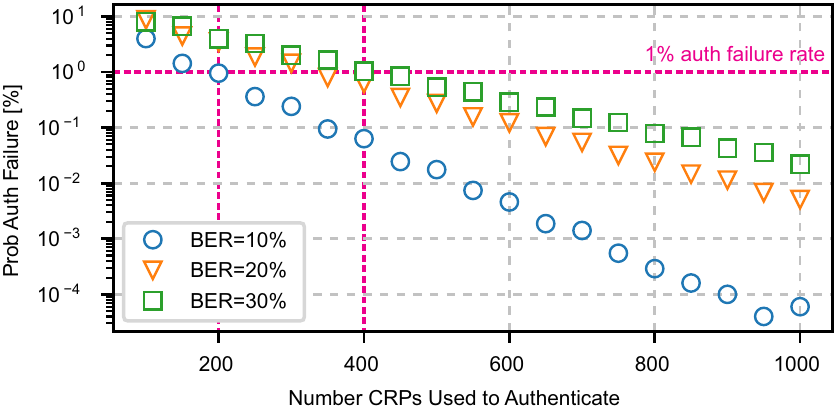}
    \caption{Probability of authentication failure simulated with 1M authentications. Uniformity of 50\% is assumed. The minimum number of correct responses is 5\% below (100\% - BER).}
    \label{fig:auth}
\end{figure}

For an IC to be deemed authentic, the number of correct responses must exceed a response threshold, otherwise the authentication fails. Response threshold is a system defined parameter which is set according to PUF response stability. If the threshold is expressed as a percentage of correct responses needed to authenticate, we can argue that it must be less than (100\% - BER), otherwise the PUF is unlikely to successfully authenticate due to noisy responses. 

As discussed in \cite{pufNmq2022}, Fig. \ref{fig:auth} simulates PUFs with different BER. Threshold was set 5\% below (100\% - BER). For example, when simulating an authentication using 200 CRPs, with a PUF that has 10\% BER, 170 correct responses are required to authenticate. As shown in Fig. \ref{fig:auth}, the probability of authentication failure falls exponentially with the number of CRPs used to authenticate. For example, for a 1\% failure rate, a PUF with 10\% BER will require 200 CRPs, while if the PUF BER is increased to 20\% or 30\%, the required CRPs to achieve the same failure rate will increase to 350, and 400, respectively.

\section{The PUF-on-PUF Architecture\label{sec:comp}}

The PUF-on-PUF (POP) architecture was proposed by Wu \etal~\cite{pufPop2019}. It utilizes composition as an alternative to increase strong PUF resilience to learning attacks. Our implementation of POP is shown in Fig. \ref{fig:arch}. It uses a two layers construction, where responses from the first layer serve as input to the second layer. Our implementation adds support for multiple first round evaluations as a low-cost alternative to increasing the number of layers. In other words, first layer responses can be reused as input challenge for additional evaluation rounds, prior to the final second layer evaluation.

Our implementation of POP uses APUFs as building block, for its simplicity, stability, and well understood security characteristics. The input challenge has 64~bits, and the number of APUF instances in the first layer matches the number of challenge bits. The first layer can be implemented with APUFs of any size, while the second layer APUF must match the number of stages with the number of APUF instances in the previous layer. In this paper, we restrict ourselves to first layer implementations where all APUFs have the same size. 

Each APUF instance uses temporal majority voting (TMV) to filter out noise, as shown in Fig. \ref{fig:tmv}. TMV performs a predetermined number of repeated evaluations, returning the most frequent response. It is important to notice that TMV is not applied to the overall composition, but to each individual APUF instance.

As described by Wu \etal, careful wiring of challenge bits in the first layer is required to avoid cryptanalysis attacks~\cite{pufPop2019}. The wiring pattern must apply each challenge bit in more than a single PUF instance. The POP wiring for a first layer implementation using 4-APUF is shown in Fig. \ref{fig:wiring}. Each $\text{4-APUF}_i$ is connected to the input challenge at offset $i$. If the sum of offset and APUF size is greater than 63, the challenge bits simply wrap around. When performing evaluations with multiple rounds, the challenge register is re-loaded with responses from APUFs in the first layer. Responses of each $\text{APUF}_i$ are used as challenge bit $i$ for first layer re-evaluation. Similarly, second layer evaluations are performed by wiring responses of each $\text{APUF}_i$ in the first layer to the stage $i$ of the second layer APUF.

\section{Testchip Design\label{sec:dsgn}}

\begin{figure}[t]
    \centering
    \includegraphics[scale=1]{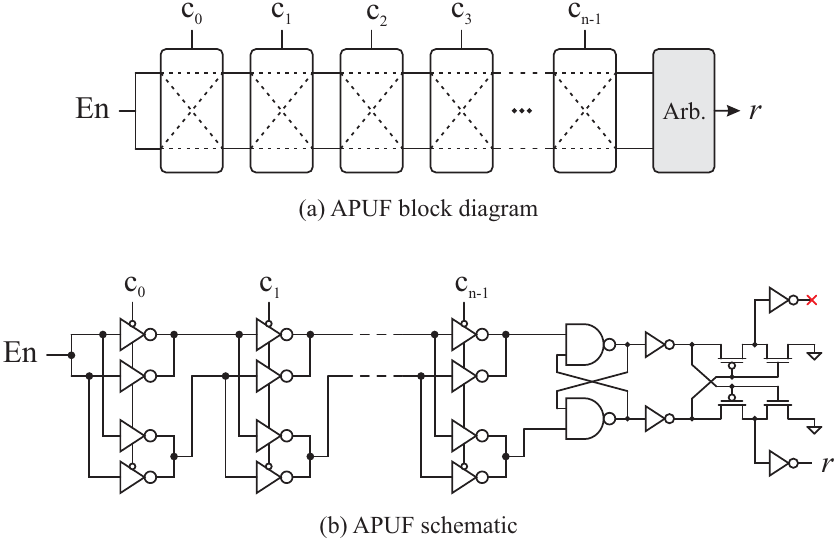}
    \caption{Arbiter PUF block diagram in (a), and implemented schematic using tri-state inverters and NAND-based arbiter in (b).}
    \label{fig:apufsch}
\end{figure}

High-level models offer a convenient alternative to test the security of new PUF architectures. The delay of each APUF stage follows a well understood normal distribution. When noise is not considered, PUF responses from high-level models tend to be indistinguishable from responses obtained from silicon. This allows designers to perform an early assessment of uniformity, uniqueness, and resilience to learning attacks. However, response stability is a key performance metric which can not be accurately estimated without silicon implementation. For this reason, we designed and implemented a testchip in 65~nm CMOS technology. Our testchip is used to evaluate the response uniformity, uniqueness, and bit error rate of our POP implementations.

\begin{figure}[t]
    \centering
    \includegraphics[scale=1]{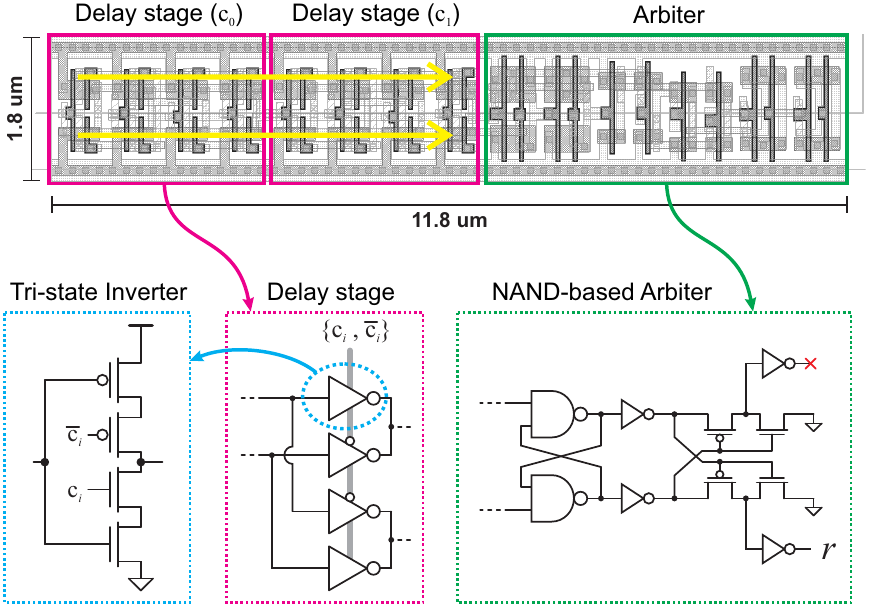}
    \caption{Custom-made layout of a 2 stages APUF. Same height as standard-cell logic for integration with automatic placement and routing tools.}
    \label{fig:layout}
\end{figure}

We use APUF as building block. Our APUF instances are designed with tri-state inverters and a NAND-based arbiter, as shown in Fig. \ref{fig:apufsch}. Layout of APUFs is custom-made to ensure identical routing of delay paths. The layout of an APUF with 2 stages is shown in Fig. \ref{fig:layout}. We designed APUFs with 2, 4, 6, 8, 12, 24, and 64 stages. Area information for each APUF is shown in Table \ref{tab:area}. Height and width dimensions are listed in $\mu m$, while area is provided in $\mu m^2$, and normalized by the NAND2 area. Our APUF cells have the same height as logic gates from the commercial standard-cell library, which allows automatic placement and routing by EDA tools. This methodology significantly reduces design effort, without loosing the performance of a custom approach.

\begin{table}[t]
    \centering
    \caption{Area cost of each APUF size.}
    \label{tab:area}
    \begin{tabular}{@{}lrrrrr@{}}
\toprule
\textbf{}   & \textbf{\# Stages} & \textbf{Height} & \textbf{Width} & \textbf{Area} & \textbf{Norm.} \\
            &                    & \textbf{($\mu m$)} & \textbf{($\mu m$)} & \textbf{($\mu m^2$)} & \textbf{(ND2)} \\ \midrule
2-APUF      & 2                  & 1.8             & 11.8           & 21.2          & 14.8                \\
4-APUF      & 4                  & 1.8             & 18.2           & 32.8          & 22.8                \\
6-APUF      & 6                  & 1.8             & 24.6           & 44.3          & 30.8                \\
8-APUF      & 8                  & 1.8             & 31             & 55.8          & 38.8                \\
12-APUF     & 12                 & 1.8             & 43.8           & 78.8          & 54.8                \\
24-APUF     & 24                 & 1.8             & 82.2           & 148.0         & 102.8               \\
64-APUF     & 64                 & 1.8             & 210.2          & 378.4         & 262.8               \\ \bottomrule
\end{tabular}
\end{table}

\begin{figure}[t]
    \centering
    \includegraphics[scale=1]{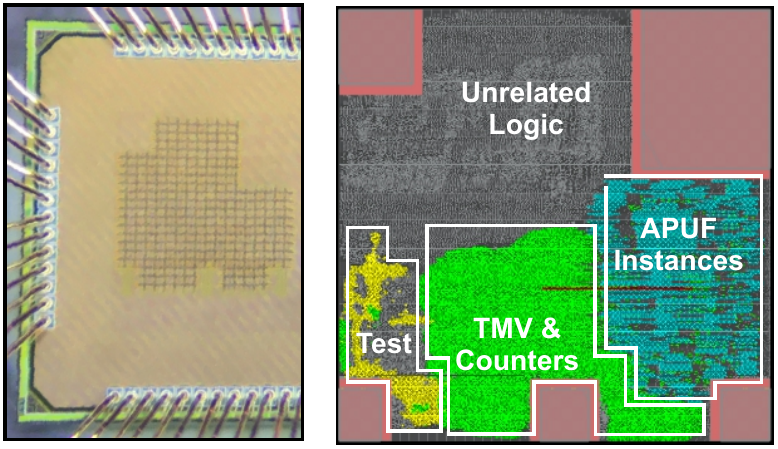}
    \caption{Die photo and layout view of implemented design. TMV logic is oversized since it uses larger counters than necessary.}
    \label{fig:chip}
\end{figure}

Die photo and layout are shown in Fig. \ref{fig:chip}. The cells highlighted in yellow implement a JTAG interface and test logic. APUF instances used in the first layer are highlighted in blue, they account for 64 instances of each APUF cell size, including 2, 4, 6, 8, 12, and 24 stages. In total, 384 APUFs are instantiated to construct 6 different first layer implementations. The PUF with 64 stages, used in the second layer, is highlighted in red. Cells highlighted in green implement the round control logic and TMV counters. The TMV logic is largely oversized for exploratory reasons, using a total of 65 counters, each with 24-bits. Results reported in section \ref{sec:meas}, show that more than 15 TMV evaluations bring diminishing returns in response stability. Therefore, the size of TMV counters may be significantly reduced. Further area optimization is possible if the first layer PUFs do not evaluate simultaneously, allowing operation with fewer TMV counters. This impacts throughput, but significantly reduces the TMV hardware size. 

In summary, each testchip includes 6 different first layer implementations. Each implementation uses a different APUF size, including 2, 4, 6, 8, 12, and 24 stages. The first layer does not mix APUFs of different sizes. There is a single 64-APUF instance, therefore, the testchip includes only one second layer implementation.

\section{Measurement Results\label{sec:meas}}

\begin{figure}[t]
    \centering
    \includegraphics[scale=1]{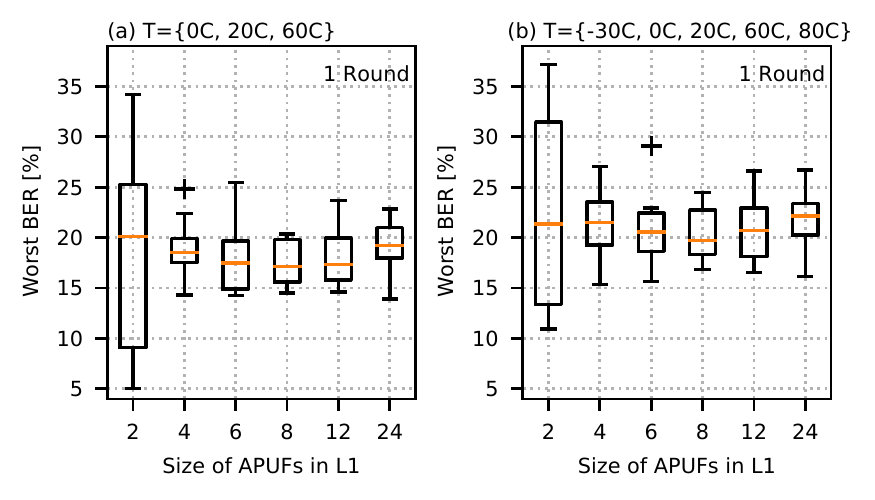}
    \caption{Worst BER for POP implementations with various APUF sizes in the first layer; in (a) for the narrow temperature set; and (b) for the wide temperature set. Measured with 5~k CRPs, single first layer evaluation (one round), and 15 TMV.}
    \label{fig:meas_rb_l1sizes}
\end{figure}

We measured a total of 10 dies to accurately assess uniformity, uniqueness, and response stability. Each die contains 6 different first layer implementations, and a single 64-APUF which implements the second layer. We performed enrollment at \temp{20}. To calculate bit error rate, CRPs are evaluated 100 times at the \textit{narrow} and \textit{wide} temperature sets, which denote \{\temp{0}, \temp{20}, \temp{60}\}, and \{\temp{-30}, \temp{0}, \temp{20}, \temp{60}, \temp{80}\}, respectively. Temporal majority voting (TMV) is used in all evaluations. TMV with a single (one) repeated evaluation is equivalent to an ordinary evaluation without temporal majority voting. Boxplots show the distribution of \textit{mean} values for different chips --- they include 10 different mean values of the respective performance metric, one from each tested chip.

Bit error rate measures the stability of PUF responses. Fig. \ref{fig:meas_rb_l1sizes} (a), and (b) plot bit error rate for POP implementations with various APUF sizes in the first layer, for the narrow and wide temperature sets, respectively. A single first layer evaluation was performed (one round). A total of 15 repeated evaluations were used for TMV. The median BER for 24-APUF-POP is 19.2\%, and 22.1\% in the narrow, and wide temperature sets, respectively. Minimum bit-error rate is found with 8-APUF-POP, where we measured median BER of 17.1\%, and 19.7\% for the narrow, and wide temperature sets. Small APUFs, with 2, 4, and 6 stages, showed a steep increase in the spread of bit error rate across different chips. This is likely related to stage bias, which is discussed in section \ref{sec:inf}.

\begin{figure}[t]
    \centering
    \includegraphics[scale=1]{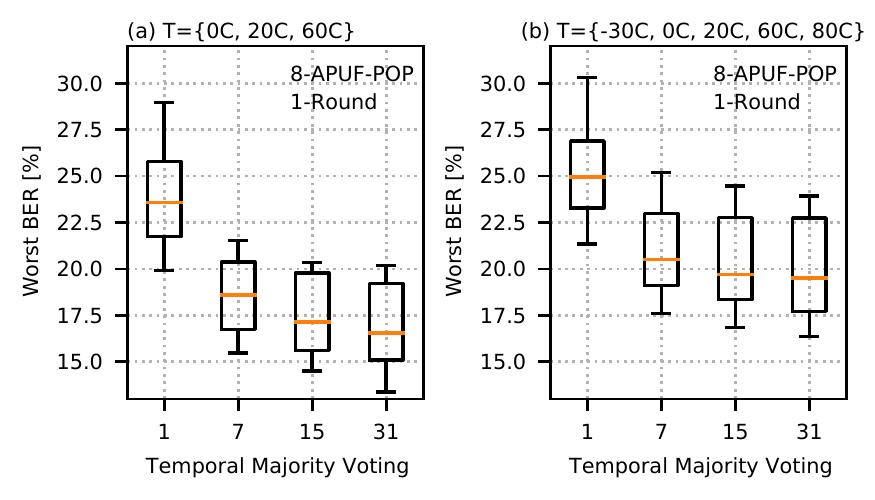}
    \caption{Worst BER for different repeated evaluation (TMV) values; (a) assess bit error rate over the narrow temperature set; and (b) over the wide temperature set. Measured with 5~k CRPs, and a single first layer evaluation (one round).}
    \label{fig:meas_rb_rnd0}
\end{figure}

\begin{figure}[t]
    \centering
    \includegraphics[scale=1]{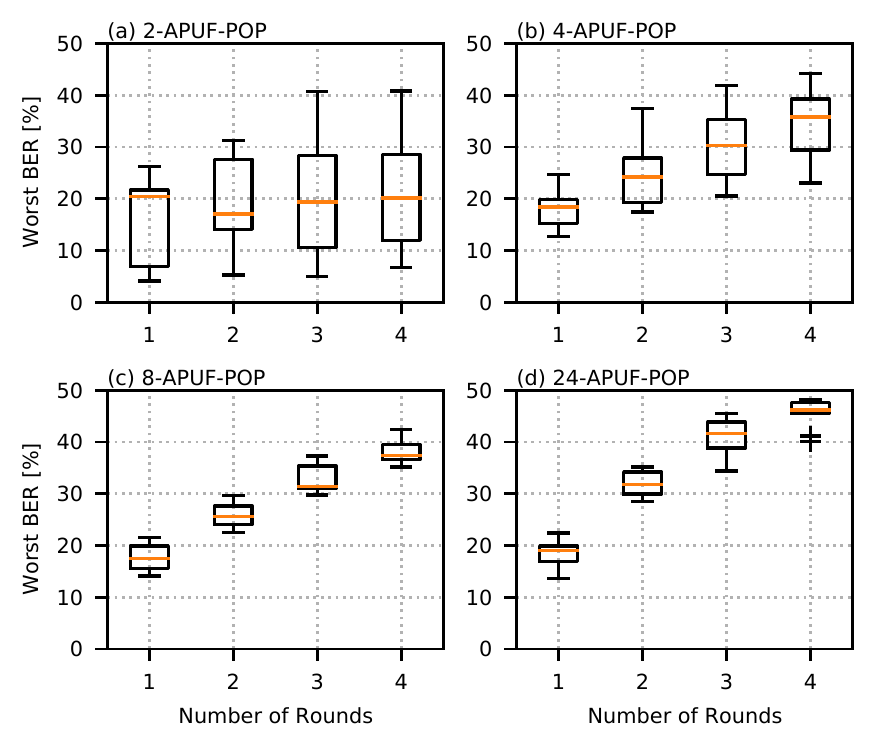}
    \caption{Bit error rate for various evaluation rounds using 2-APUF-POP in (a), 4-APUF-POP in (b), 8-APUF-POP in (c), and 24-APUF-POP in (d). Measured in the narrow temperature set, with 1~k CRPs, and 15 TMV.}
    \label{fig:meas_rb}
\end{figure}

Temporal majority voting (TMV) performs multiple evaluations of a PUF instance to remove noise from responses. We assessed the impact of TMV in POP response stability. Results plotted in Fig. \ref{fig:meas_rb_rnd0} show bit error rate for different TMV settings. In (a), the median BER for 8-APUF-POP, in the narrow temperature set, is 23.5\%, 18.5\%, 17.1\%, and 16.5\%, for 1, 7, 15, and 31 repeated TMV evaluations. Increasing the number of repeated evaluations quickly reaches diminishing returns. A similar trend was observed for other sizes of POP, in both temperature sets. Therefore, 15 TMV offers a reasonable compromise between throughput and bit error rate for composite evaluations.

Our POP implementations adds support to multiple first layer evaluation rounds. We assessed the response stability for 1, 2, 3, and 4 evaluation rounds, using 1~k CRPs. Results are shown in Fig. \ref{fig:meas_rb} (a), (b), (c), and (d) for 2, 4, 8, and 24 stages in first layer APUFs, respectively. All implementations showed median BER near 20\% for single round evaluations. Larger APUFs in the first layer lead to lower response stability when using additional rounds, as expected. Similarly to Fig. \ref{fig:meas_rb_l1sizes}, small APUFs in the first layer show a wide spread of BER measurement across multiple chips. Therefore, median BER for 2-APUF-POP, and 4-APUF-POP, may be a misleading performance metric, if not accompanied by the corresponding standard deviation. 

\begin{figure}[t]
    \centering
    \includegraphics[scale=1]{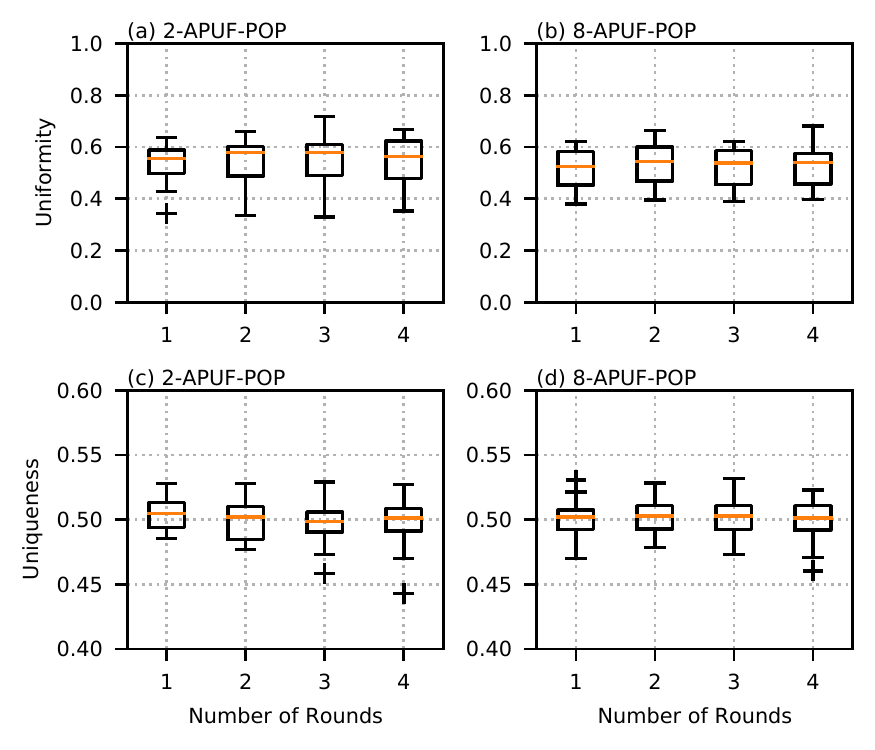}
    \caption{Uniformity and uniqueness for various evaluation rounds for 2-APUF-POP in (a, c), and 8-APUF-POP in (b, d). Measurements were performed with 1~k CRPs at \temp{20}, with 15 TMV.}
    \label{fig:meas_uu}
\end{figure}

We also measured uniformity and uniqueness of POP instances in our testchip. The former, captures the balance of zeros and ones in the final response, while the later, measures the normalized hamming distance of different instances. Measured uniformity and uniqueness for various rounds are plotted in Fig. \ref{fig:meas_uu}, for 2-APUF-POP in (a, c), and 8-APUF-POP in (b, d). Both uniformity, and uniqueness showed a small variation in performance values across different sizes of POP. The median uniformity for 2-APUF-POP, and 8-APUF-POP, with one evaluation round, is 0.55, and 0.52, respectively. The corresponding median uniqueness is 0.50 for both 2-APUF-POP and 8-APUF-POP. Other POP implementations with different sizes presented similar results. 

In section \ref{sec:chlhd} we show that small APUFs in the first layer lead to poor hamming distance in the intermediate responses. Such result is not noticeable when evaluating the final output of POP, but it is likely the cause of security vulnerabilities studied in the next section.

\section{Security Assessment \label{sec:sec}}

The primary motivation for composite strong PUFs is to increase security against learning attacks. We start this section discussing the applicability of logistic regression, cryptanalysis, and reliability attacks to POP. Next we describe how deep neural networks (DNNs) are used to model strong PUFs. We assess the learning resilience of our POP implementations against DNNs using up to 10~M CRPs. Next, we discuss influential bits, and hamming distance of intermediate responses, showing that compositions do not necessarily preserve the security properties of PUFs --- the size of composing PUFs plays a crucial role.

\subsection{Learning Attack Results\label{sec:att}}

\begin{table}[t]
    \centering
    \caption{Learning attack results.}
    \label{tab:dnn}
    \begin{threeparttable}
        \begin{tabular}{@{}lrrrrr@{}}
\toprule
\textbf{Implementation}     & \textbf{Rounds} & \textbf{BER}        & \textbf{Area (ND2)}            & \textbf{Accuracy}     \\ \midrule
2-APUF-POP           & 1                  & 20.0\%                  & 1.2 k                          & 81.8\%                \\
4-APUF-POP           & 1                  & 18.5\%                  & 1.7 k                          & 74.7\%                \\
6-APUF-POP           & 1                  & 17.4\%                  & 2.2 k                          & 48.0\%                \\
8-APUF-POP           & 1                  & 17.1\%                  & 2.7 k                          & 51.5\%                \\
12-APUF-POP          & 1                  & 17.3\%                  & 5.4 k                          & 49.6\%                \\
24-APUF-POP          & 1                  & 19.2\%                  & 6.8 k                          & 54.4\%                \\ \midrule
2-APUF-POP           & 1                  & 20.0\%                  & 1.2 k                          & 81.8\%                \\
2-APUF-POP           & 2                  & 17.0\%                  & 1.2 k                          & 91.3\%                \\
2-APUF-POP           & 4                  & 20.1\%                  & 1.2 k                          & 96.1\%                \\
2-APUF-POP           & 8                  & --                      & 1.2 k                          & 99.5\%                \\ \midrule
24-APUF-POP          & 1                  & 19.0\%                  & 6.8 k                          & 54.4\%                \\
24-APUF-POP          & 2                  & 31.7\%                  & 6.8 k                          & 55.0\%                \\
24-APUF-POP          & 4                  & 41.6\%                  & 6.8 k                          & 56.4\%                \\
24-APUF-POP          & 8                  & --                      & 6.8 k                          & 56.2\%                \\
\bottomrule
\end{tabular}

        \begin{tablenotes}[para,flushleft]
            Notes: worst BER is the worst bit error rate (median) from \{\temp{0}, \temp{20}, \temp{60}\}, using 15 repeated evaluations for TMV. The area includes only APUF instances (control and TMV logic not included). All attacks were performed using deep neural networks (DNNs), with 10~M CRPs and 72~h of training time.
        \end{tablenotes}
    \end{threeparttable}
\end{table}

\subsubsection{Logistic regression and cryptanalysis}

logistic regression (LR), and cryptanalysis attacks are described in ~\cite{seminalAttack2013} and ~\cite{attackCryptanalysis2015}. LR uses gradient descent to find coefficients of a linear model that minimizes the prediction error. Modeling POP with LR is non-trivial, since careful mathematical manipulation is required for an adequate linear fitting~\cite{pufPop2019}. The cryptanalysis attack exploits the mapping of challenge bits to different APUFs in the first layer. It was shown in~\cite{pufPop2019}, that such attacks are ineffective against POP due to the wiring scheme used in the first layer --- every challenge bit is fed to more than a single first layer PUF. 

\subsubsection{Reliability based attacks}

reliability based attacks were initially introduced in~\cite{attackRelb2015}. The key insight is that CRPs with small delay difference are more susceptible to noise. Authors demonstrated the attack against XOR-APUFs using response stability as side-channel information. The POP architecture, however, uses a construction where noisy CRPs in the first layer affect the final output with different probabilities. Such characteristic is described in detail in section \ref{sec:prout}. Hence, analogously the the formal proof provided for the interpose PUF in~\cite{pufInterpose2019}, reliability-based attacks are unlikely to obtain better prediction accuracy than other learning attacks that do not exploit response stability information.

\subsubsection{Deep neural networks}

deep neural networks (DNNs) are emerging as an efficient attack technique capable of learning complex PUF structures. DNNs do not require a mathematical model of the PUF being modelled. We use a 12-layer DNN architecture proposed in~\cite{attackDNN2019} for our DNN attacks. The input and output layers have 64, and 2 units, respectively. Hidden layers have 2000 units. Our attack experiments use CRPs from a high-level model to avoid noise as confounding factor for prediction accuracy. Table \ref{tab:dnn} summarizes DNN attack results using 10~M CRPs.  The \textit{Implementation} column specifies POP parameters used, for example, the 2-APUF-POP uses 64 instances of 2-APUF in the first layer. All our POP implementations use a 64-APUF instance in the second layer. The \textit{Rounds} column refers to the number of first layer evaluations used, prior to the second layer evaluation. The \textit{BER} column reports the median worst bit error rate among the temperature points of \{\temp{0}, \temp{20}, \temp{60}\}. The \textit{Area} column reports the silicon area (normalized by the NAND2 area) for all APUF instances, excluding control, and TMV logic. The \textit{Accuracy} column reports the accuracy obtained after 72~h of training using a Quadro P4000 GPU. 

Results reported in Table \ref{tab:dnn} show that the DNN model achieved accuracy of 81.8\% for 2-APUF-POP using 1 round. This implementation has median BER of 20.0\%, and uses an area of 1.2~k ND2. Increasing the size of APUFs in the first layer to 4, 6, 8, 12, and 24 stages reduces prediction accuracy to 74.7\%, 48.0\%, 51.5\%, 49.6\%, and 54.4\%, respectively. Therefore, our results suggest that increasing the size of APUFs in the first layer strengthens the POP composition, at an area and response stability cost.

We explored multiple first layer evaluation rounds as a cost-effective approach for strengthening POP against learning attacks. Table \ref{tab:dnn} reports bit error rate and prediction accuracy result for 2-APUF-POP and 24-APUF-POP, using 1, 2, 4, and 8 rounds. Response stability falls as more evaluation rounds are used. We measured median BER of 17.7\% and 31.7\% for 2-APUF-POP and 24-APUF-POP when using 2 evaluation rounds, respectively. Prediction accuracy for 24-APUF-POP did not shown significant change when varying number of rounds, however, prediction accuracy for 2-APUF-POP increased when more rounds are used. This result was unexpected, and counter-intuitive. We carefully reflect on possible explanations for such outcomes in the next sections.

\subsection{Probability of Output Change\label{sec:prout}}

The DNN prediction accuracy when using small APUFs in the first layer, reported in section \ref{sec:att}, deserves additional investigation. In this section, we use the concept of strict avalanche criterion (SAC) to look into the differences between APUFs of various sizes.

\begin{figure}[t]
    \centering
    \includegraphics[scale=1]{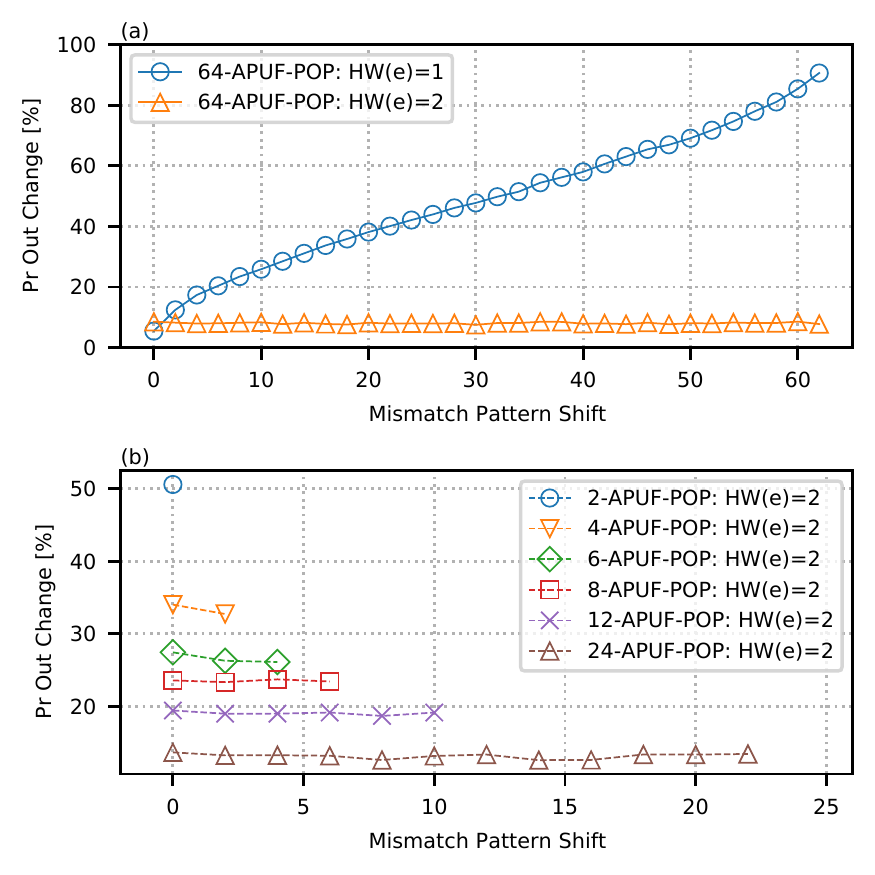}
    \caption{Simulated probability of output change when evaluating a random challenge, before, and after it is XORed with a mismatch pattern. Results in (a) for 64-APUF, and in (b) for smaller APUF sizes. The HW(\textbf{e})=2 only includes mismatch patterns where nonzero bits are adjacent.}
    \label{fig:prout_apuf}
\end{figure}

As defined in~\cite{othersSac1985}, if a cryptographic function is to satisfy the strict avalanche criterion (SAC), then, each output bit should change with a probability of one half, whenever a single input bit is complemented. In~\cite{othersSacPufFirst2008, othersSacPuf2016} this concept was extended to strong PUFs, where authors measure the probability of output response change given a single bit change in the input challenge. Furthermore, it was demonstrated that the probability of output change for the APUF, depends on the distance between toggled bit and the arbiter. Challenge bits applied to stages near the arbiter are more likely to cause a change in the response. This result is reproduced in Fig. \ref{fig:prout_apuf} (a). Using simulation, we estimate the probability of output change for 64-APUF when evaluating a random challenge, before, and after it is XORed with a mismatch pattern \textbf{e}. When $\text{HW}(\vect{e})=1$, a single challenge bit will toggle between evaluations. In the plots of Fig. \ref{fig:prout_apuf}, the position of the toggled bit is shifted towards the arbiter, and is denoted as mismatch pattern shift. When the pattern shift is zero, probability of output change is 5.5\%, but as the toggled bit nears the arbiter, probability of output change increases, reaching 90.5\% at the last APUF stage. This result can be intuitively explained by the wire permutation present in every stage of the APUF, and the cumulative nature of the delay path.

We also estimate the probability of output change when two consecutive challenge bits are toggled. This is plotted in Fig. \ref{fig:prout_apuf} (a) as $\text{HW}(\vect{e})=2$, showing that, for 64-APUF, the probability of output change remains nearly constant, at 8\%. This result represents a more realistic view on the SAC criterion for APUFs, where minimal input change requires toggling two adjacent stages, instead of a single one. The same technique was applied to APUF of various sizes in Fig. \ref{fig:prout_apuf} (b). The estimated probability of output change for APUFs with 24, 12, 8, 6, 4, and 2 was 13.2\%, 19\%, 23.6\%, 26.5\%, 33.5\%, and 50.5\%, respectively. The increase in probability of output change for smaller sizes of APUFs gives an important insight to understand the results found in section \ref{sec:att}: the influence of individual stages on the output increases, as APUF size decreases.


\subsection{Influential Bits and Stage Bias\label{sec:inf}}

\begin{algorithm}[t]
    \caption{Stage bias assessment for an APUF instance.}
    \label{algo:inf}
    \begin{algorithmic}[1]
\STATE \textbf{let} $NC$ be the number of challenges
\STATE \textbf{let} $CW$ be the challenge width in bits
\STATE \textbf{let} $y$ and $n$ be zero initialized matrices of size $(2, CW)$
\FOR{$i = 0$ to $NC - 1$}
    \STATE $c = \text{RandomizeChallenge}()$
    \STATE $r = \text{EvaluateResponse}(c)$
    \STATE $p = 0$
    \FOR{$j = 0$ to $CW - 1$}
        \STATE $p = p \oplus c[j]$
    \ENDFOR
    \FOR{$j = 0$ to $CW - 1$}
        \STATE $t = c[j]$
        \STATE $p = p \oplus t$
        \STATE $y[t,j] \pluseq  (r \oplus p)$
        \STATE $n[t,j] \pluseq 1$
    \ENDFOR
\ENDFOR
\STATE $y = y / n$
\end{algorithmic}
\end{algorithm}


Influential bits were previously studied in~\cite{attackInfbitsSingleLayer2014, attackInfbitsDiffLin2014, attackInfbits2016}. Authors showed how distinct challenge bits have different influence on the output of a bistable ring PUF (BR-PUF). Based on the value of a few influential challenge bits, it is possible to predict responses with high accuracy~\cite{attackInfbits2016}. To avoid confusion with previous work nomenclature, we denote the \textit{influence} of each challenge bit as \textit{stage bias}. This section performs an assessment of stage bias in APUFs of various sizes, leading to conclusions that help explain learning attack results obtained in section \ref{sec:att}.

To the best of our knowledge, no previous literature reports the measurement of stage bias in APUFs. In~\cite{attackInfbits2016}, authors used an algorithm described in~\cite{othersInfbitsAlgo2012} to assess stage bias of BR-PUFs, however, the algorithm is only suitable for monotone Boolean functions. Therefore, we introduce Algorithm \ref{algo:inf} for measuring stage bias in APUFs. The main idea is to evaluate a set of randomized challenges, keeping track of responses statistics per stage, and per challenge bit value. The key insight of Algorithm \ref{algo:inf} is on line 14. When summing the response, $r$, for challenge bit value $t$, at stage $j$, the response is conditionally inverted (XORed) with $p$, where $p$ is a \textit{parity bit} (reduced XOR operation) over the challenge bits from position $j+1$, onwards. If the number of twisted stages after position $j$ is odd, the value of $p$ will be 1, which then inverts the response $r$. The final stage bias is stored in the matrix $y$, indexed by challenge bit value, and by stage position. Notice that the division operation in line 18 is element-wise.

\begin{figure*}[t]
    \centering
    \includegraphics[scale=1]{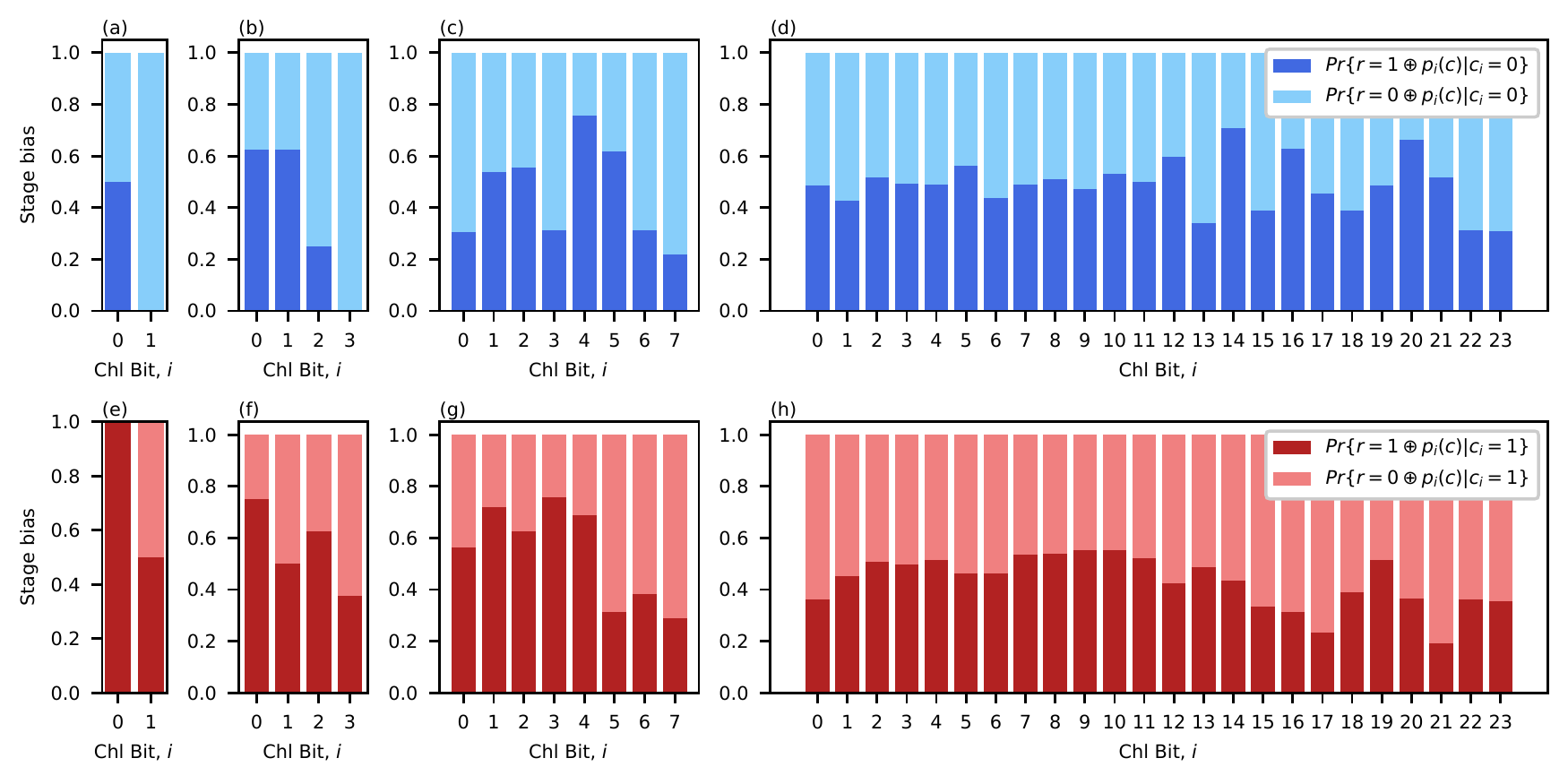}
    \caption{Stage bias calculated using CRPs from our testchip. The first row shows biases when $c_i=0$, while the second row shows biases when $c_i=1$. The plots (a, e), (b, f), (c, g), and (d, h), refer to APUFs with 2, 4, 8, and 24 stages, respectively. We used 100~k CRPs to calculate stage bias for the 24-APUF, while APUFs of 2, 4, and 8 stages were enumerated (all CRPs were collected).}
    \label{fig:meas_infbits}
\end{figure*}

Using CRPs from our testchip, we calculated stage bias for different APUF sizes. The results are shown in Fig. \ref{fig:meas_infbits}. A distinction was made for stage bias when $c_i$ is zero (first row), and one (second row), since challenge bits select between two pairs of inverters in each stage, and each pair exerts different influence on the response. The plots (a, e), (b, f), (c, g), and (d, h), refer to APUFs with 2, 4, 8, and 24 stages, respectively. The results refer to a single APUF instance of each size (not the overall POP composition).

Results plotted in Fig. \ref{fig:meas_infbits} express the probability of response $r$ being equal to $(1 \oplus p_i(\vect{c}))$, given challenge bit $c_i$ is zero, or one. The term $p_i(\vect{c})$ is denoted as \textit{parity}. It will have a value of one when the challenge imposes an odd number of wire twists between the stage under analysis $i$, and the arbiter. Parity is calculated as

\begin{equation}
p_i(\vect{c}) = \bigoplus_{j=i+1}^{n-1} c_j.\label{eq:p}
\end{equation}

For example, based on data from Fig. \ref{fig:meas_infbits} (f) for the 4-APUF, the probability of $r=1$, given $c_3=1$, is 0.38, or equivalently, probability of $r=0$, given $c_3=1$, is 0.62. In this case, the parity calculated by Eq. \ref{eq:p} is zero, since there are no stages that could twist the wires between position 3 and the arbiter. As an alternative example, the stage bias reported in Fig. \ref{fig:meas_infbits} (c) for the 8-APUF shows that, the probability of $r=(1 \oplus p_i(\vect{c}))$, given $c_4=0$, is 0.76. Therefore, all challenges which have $c_4=0$, and an even number of ones in $(c_5, c_6, c_7)$, have 0.76 probability of evaluating to 1. Moreover, challenges that have $c_4=0$, but an odd number of ones in $(c_5, c_6, c_7)$, have a 0.76 probability of evaluating to 0.

\begin{figure}[t]
    \centering
    \includegraphics[scale=1]{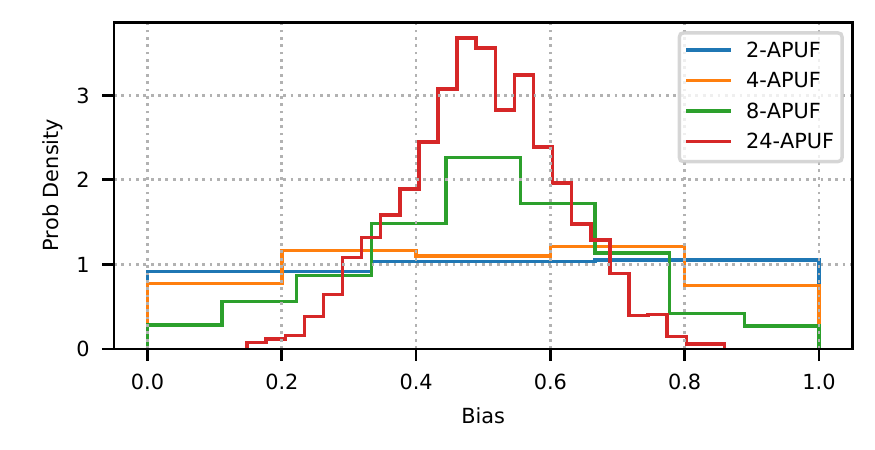}
    \caption{Distribution of stage bias obtained from simulation with 100 APUF instances and 3~k CRPs.}
    \label{fig:infbits_hist}
\end{figure}

Large stage bias deviations from 0.5 are undesirable, since they grant certain challenge bits an unfair influence over the response. It was also shown that large stage bias can be exploited by attackers~\cite{attackInfbits2016}. To understand how stage bias varies across APUFs of different sizes, we simulated 100 APUF instances using 3~k CRPs, and plotted the stage bias distribution in Fig. \ref{fig:infbits_hist}. The stage bias mean is 0.5 for all APUF sizes, but the standard deviation for 24, 8, 4, and 2-APUF are 0.12, 0.20, 0.29, and 0.40, respectively. Therefore, we may conclude what is also apparent in the measurements presented in Fig. \ref{fig:meas_infbits}, fewer APUF stages increase the likelihood of large stage bias deviations.

\subsection{Hamming Distance of Intermediate Responses\label{sec:chlhd}}

Previous section assessed the effects of smaller APUFs on stage bias, showing that reducing the size of APUFs creates challenge bits that hold large influence over the final response. Such result motivates an investigation of the hamming distance between responses produced by the first layer of the POP architecture, over multiple rounds (same challenge), and across multiple challenges. Although mathematically similar, we avoid using the term uniqueness for such experiment, since it is not applied to the final POP response.

\begin{figure}[t]
    \centering
    \includegraphics[scale=1]{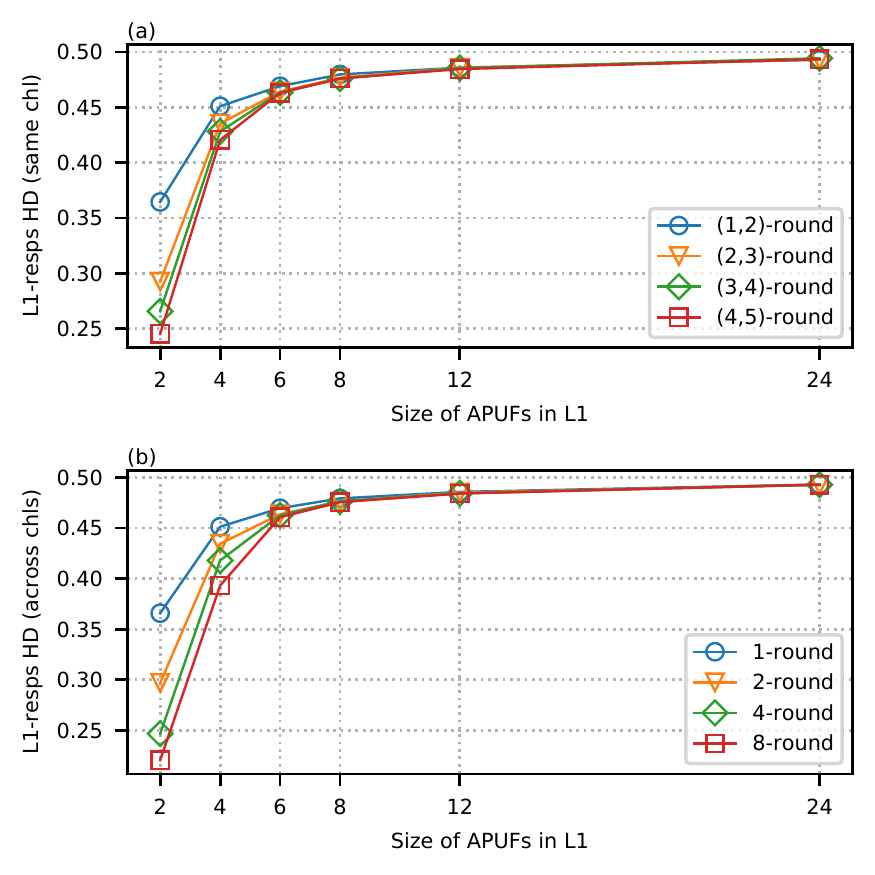}
    \caption{Simulation of the average normalized hamming distance (HD) between responses for various sizes of APUFs in first layer (L1). Across multiple rounds, for the same challenge in (a); and across multiple challenges, after 1, 2, 4, and 8 evaluation rounds in (b).}
    \label{fig:chlhd}
\end{figure}

The normalized hamming distance (HD) measures the distance between two numbers, divided by their length. The normalized HD is herein denoted as distance, for short. For example, if the distance between two numbers is 0.5, it implies that half the bits of their binary representation differ. Fig. \ref{fig:chlhd} (a) plots the average distance between the responses produced by the POP first layer, across multiple rounds, for the same challenge, for various APUF sizes. For instance, the data denoted as \textit{(1,2)-round} reports the average distance between responses produced from first to the second evaluation round. While implementations with 24-APUF show nearly ideal distance of 0.5 across all evaluation rounds, reducing the size of APUFs gradually degrades the distance between responses. In implementations with 2-APUF, the average distance between responses from first to the second round is 0.36, which implies that 64\% of response bits from the first evaluation remained unchanged after the second evaluation. As the number of rounds increases, average distance for 2-APUF continues to fall, reaching 0.24 for responses between fourth, and fifth rounds.

We also examine the average distance of first layer responses, across multiple challenges, after 1, 2, 4, and 8 evaluation rounds. Results are plotted in Fig. \ref{fig:chlhd} (b). Implementations using 24-APUF show nearly ideal distance of 0.5 across multiple challenges, but reducing the size of APUFs, gradually degrades the distance between responses --- this time, across multiple challenges. For example, 2-APUF implementations show average distance of 0.36, 0.30, 0.25, and 0.22 when evaluated with 1, 2, 4, and 8 rounds. Moreover, the data suggests that even with a single evaluation round, small APUFs fail to produce responses with distance near 0.5. Another perspective to such result, is to consider that small APUFs limit the challenge space of the second layer APUF, seriously impacting to the learning resilience of the overall composition.

The poor hamming distance performance of smaller APUFs, likely caused by larger stage bias, is a plausible cause for the learning results observed in section \ref{sec:att}. The results in Fig. \ref{fig:chlhd} suggest a strong performance drop for 2 and 4 stages APUFs, which agrees with our attack results, where DNNs failed to obtain generalized knowledge for 6-APUF-POP implementations and above (see section \ref{sec:att}). It is also important to notice that our results do not assess the benefits of multiple round evaluations for larger APUFs in the first layer. In terms of prediction accuracy, those implementations were already resilient to DNN attacks with a single round. Our analysis shows, however, that there is no apparent reduction in challenge space when multiple evaluations rounds are used with first layer implementation of 12, and 24 stages.

\section{Conclusion\label{sec:conc}}

We explored the design space of the POP architecture using APUFs of various sizes. We performed extensive DNN attacks to assess the security of POP. Our results endorse POP resilience to learning attacks when using APUFs with 6, or more, stages in the first layer. Compositions using APUFs with 2, and 4 stages are shown vulnerable to DNN attacks. Moreover, POP implementations with 2 stage APUFs in the first layer show a trend of higher prediction accuracy as the number of evaluation rounds increases. To study such result, we extended previous techniques of influential bits to assess stage bias in APUF instances. Our data suggests that small APUFs in the first layer limit the challenge space of the second layer APUF, showing that compositions not always preserve security properties of PUFs. Measurements from our testchip show that minimum bit error rate is obtained when using APUFs with 8 stages, while fewer APUF stages lead to a large spread of bit error rate across different chips.

\bibliographystyle{plain}
\bibliography{pop}

\vspace{-0.4in}
\begin{IEEEbiography}[{\includegraphics[width=1in,height=1.25in,clip,keepaspectratio]{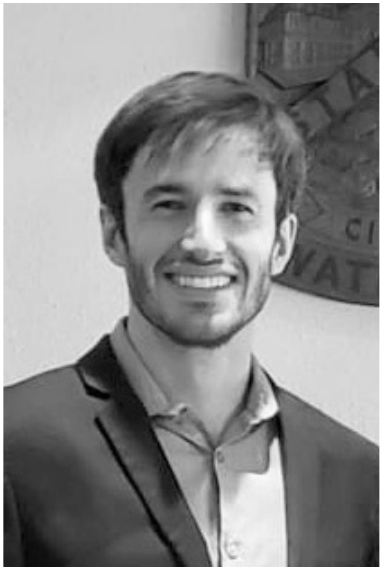}}]{Kleber Stangherlin}
Kleber received his B.Sc. in Electrical Engineering at PUCRS in Brazil. He obtained his M.Sc. in Microelectronics at UFRGS, also in Brazil. He has more than 6 years of industry experience designing security focused integrated circuits. He had key contributions to the cryptographic cores and countermeasures used in the first EAL 4+ certified chip designed in the southern hemisphere. Currently, Kleber is pursuing a PhD at University of Waterloo in Canada, where he conducts research in hardware security.
\end{IEEEbiography}

\vfill
\newpage

\vspace{-0.4in}
\begin{IEEEbiography}[{\includegraphics[width=1in,height=1.25in,clip,keepaspectratio]{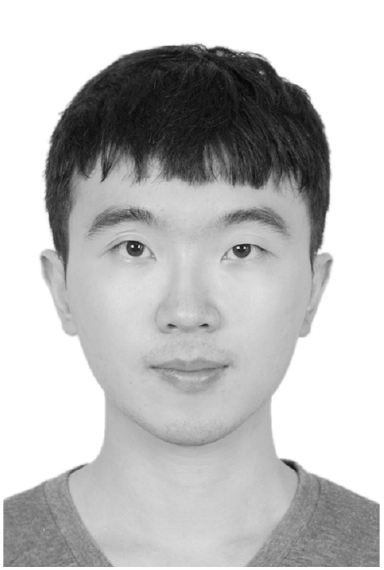}}]{Zhuanhao Wu}
Zhuanhao Wu (S'19) received the B.S. degree in computer science and technology from Nankai University, Tianjin, China in 2017, and the MASc degree in electrical and computer engineering from University of Waterloo, Ontario, Canada, in 2019, where he is currently pursuing the Ph.D. degree with the department of Electrical and Computer Engineering. His current research interests include hardware security, machine learning, and computer architecture.
\end{IEEEbiography}

\vspace{-0.4in}
\begin{IEEEbiography}[{\includegraphics[width=1in,height=1.25in,clip,keepaspectratio]{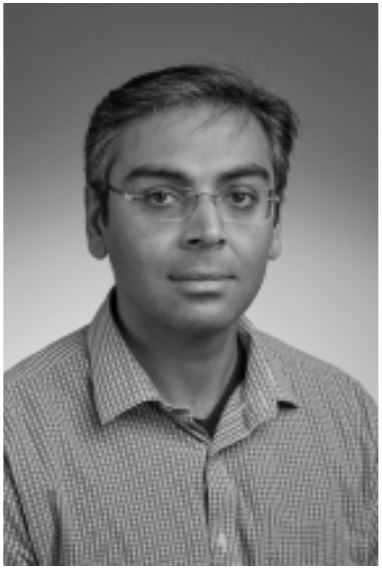}}]{Hiren Patel}
Hiren Patel is a Professor in the Department of Electrical and Computer Engineering at the University of Waterloo. Prior to the University of Waterloo, Hiren was a postdoctoral fellow at the University of California, Berkeley working in the Ptolemy group with Edward A. Lee. His research is in the design, analysis, and implementation of computer hardware and software. Currently, his research areas of interest are in real-time embedded systems, computer architecture, hardware architectures for machine learning and artificial intelligence, and security.
\end{IEEEbiography}

\vspace{-0.4in}
\begin{IEEEbiography}[{\includegraphics[width=1in,height=1.25in,clip,keepaspectratio]{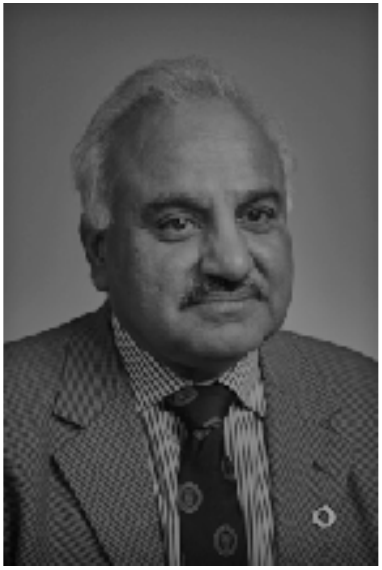}}]{Manoj Sachdev}
Manoj Sachdev is a Professor in the Department of Electrical and Computer Engineering at the University of Waterloo. He has contributed to over 225 conference and journal publications, and has written 5 books. He also holds more than 30 granted US patents. Along with his students and colleagues, he has received several international research awards. He is a Fellow of the Institute of Electrical and Electronics Engineers (IEEE), and Fellow of the Engineering Institute of Canada. Professor Sachdev serves on the editorial board of the Journal of Electronic Testing: Theory and Applications. He is also a member of program committee of IEEE Design and Test in Europe conference.
\end{IEEEbiography}

\vfill

\end{document}